\documentclass[twocolumn,iop]{emulateapj}

\usepackage{amsmath}
\usepackage{amssymb}
\usepackage{hyperref}
\usepackage{graphics,color}

\newcommand\jcap{\ref@jnl{J. Cosmology Astropart. Phys.}}

\newcommand{\hMpc}{$h^{-1}\,{\rm Mpc}$}
\newcommand{\hkpc}{$h^{-1}\,{\rm kpc}$}
\newcommand{\degsq}{\,deg$^{2}$}

\newcommand{\kms}{\,km\,s$^{-1}$}

\newcommand{\qsopair}{CFHTLS J0221-0342}
\newcommand{\qsoa}{QSO-A}
\newcommand{\qsob}{QSO-B}
\newcommand{\sextractor}{SExtractor}
\newcommand{\psfex}{PSFEx}
\newcommand{\cfhtlsurl}{\url{http://www.cadc-ccda.hia-iha.nrc-cnrc.gc.ca/en/megapipe/cfhtls/index.html}}
\newcommand{\xmmxxlurl}{\url{http://www.sdss.org/dr12/algorithms/ancillary/boss/xmmfollowup/}}

\begin{document}

\title{A Constraint on Quasar Clustering at $z=5$ from a Binary Quasar}

\slugcomment{Accepted for publication by AJ}

\thanks{Observations reported here were obtained at the MMT Observatory, a joint facility of the Smithsonian Institution and the University of Arizona.}

\author{
Ian~D.~McGreer\altaffilmark{1},
Sarah~Eftekharzadeh\altaffilmark{2},
Adam~D.~Myers\altaffilmark{2}, and
Xiaohui~Fan\altaffilmark{1}
}

\altaffiltext{1}{
Steward Observatory, 933 North Cherry Avenue, Tucson, AZ 85721, USA.
}
\email{imcgreer@as.arizona.edu}

\altaffiltext{2}{
Department of Physics and Astronomy, 
University of Wyoming, 
Laramie, WY 82071, USA.
}

\shortauthors{McGreer et al.}
\shorttitle{A $z=5$ Binary Quasar}

\begin{abstract}
We report the discovery of a quasar pair at $z=5$ separated by 21\arcsec.
Both objects were identified as quasar candidates using simple color selection
techniques applied to photometric catalogs from the CFHT Legacy Survey 
(CFHTLS). Spectra obtained with the MMT present no discernible offset in 
redshift between the two objects; on the other hand, there are clear 
differences in the emission line profiles and in the multiwavelength spectral 
energy distributions that strongly disfavor the hypothesis that they are
gravitationally lensed images of a single quasar. Both quasars are 
surprisingly bright given their proximity (a projected separation of 
$\sim135$~kpc), with $i=19.4$ and $i=21.4$. Previous measurements of the 
luminosity function demonstrate that luminous quasars are extremely rare at 
$z=5$; the existence of this pair suggests that quasars have strong 
small-scale clustering at high redshift. Assuming a real-space correlation 
function of the form $\xi(r) \propto (r/r_0)^{-2}$, this discovery implies a 
correlation length $r_0 \ga 20$\hMpc, consistent with a rapid strengthening 
of quasar clustering at high redshift as seen in previous observations and 
predicted by theoretical models where feedback effects are inefficient at 
shutting down black hole growth at high redshift.
\end{abstract}

\section{Introduction}
\label{sec:introduction}

The discovery that luminous quasars cluster strongly at redshifts approaching 
$z\sim 4$ \citep[with a scale length of $r_0\sim25$\,\hMpc;][]{She07} 
potentially poses an interesting cosmological challenge. It could be that 
quasar clustering strongly declines with luminosity at high redshift, meaning
that current samples only trace the most strongly clustered sources. But, 
faint quasars do not appear to cluster significantly more weakly than bright 
quasars at $z\sim2.5$ \citep{Whi12,Eft15} or below 
\citep[e.g.][]{daA08,She09,She13}. Thus, if quasar clustering is highly 
luminosity-dependent at $z\sim4.5$, then quasars (as a population) would have 
to alter rapidly over 10\% of the Hubble Time, then change more quiescently
over the final 80\% of cosmic history. Further, most models invoke a narrow 
range of halo mass for a wide range of quasar luminosity in order to 
reproduce the quasar luminosity function (e.g., \citealt{Lid06}; see also 
the discussion in Appendix B of \citealt{Whi12}).

Alternatively, quasars at high redshift could simply trace the growth of 
their parent dark matter halos while they are actively accreting. 
Such a scenario essentially represents the maximal possible increase in 
clustering amplitude with redshift; under scenarios other than this ``maximal 
growth'' model the correlation length of quasars should eventually diminish
at high redshift. Fig.\ 13 of \citet{Hop07} illustrates this point --- 
quasar clustering should decrease at $z>4$ for scenarios in which quasars 
are efficiently quenched.

Observations of quasar clustering at $z>4$ are currently limited to 
relatively small samples of highly luminous quasars \citep{She07}. Improving
this situation is a significant challenge, given that it requires expensive
spectroscopic campaigns targeting faint candidates at low sky density
($\sim$1~deg$^{-2}$). One promising avenue is the study of pairs of quasars 
that are separated by both a small angle on the plane of the sky and by a 
small velocity window in redshift space. Such pairs, often called ``binary 
quasars'' in the literature, are sufficiently rare to confirm with dedicated 
spectroscopic follow-up, but have a very strong clustering strength. Binary 
quasars can therefore be used to estimate the correlation length of quasar
clustering even using small samples \citep[e.g.][]{Hen06,Mye08,She10}. Of 
order a dozen $z\sim4$ binary quasars with proper separations of less than 
$\sim1$~Mpc are currently known \citep[e.g.][]{Hen10}.

The highest redshift binary quasar discovered to date is a quasar pair at 
$z=4.26$ separated by 33\arcsec, or about 230\,kpc proper, on the plane of 
the sky \citep{Sch00}. The pair was discovered serendipitously---while 
spectroscopically confirming an $i=20.4$ quasar candidate a second $i=21.3$ 
quasar at the same redshift happened to be located in the slit. This single 
quasar pair was sufficient to ascertain that $z \sim4.25$ quasars cluster with 
a correlation length of $r_0\sim10$--30\,Mpc, an observation later confirmed 
using much larger samples by \citet{She07}. In this paper, we present a 
similar find. During a survey of $4.7~\la\,z~\la\,5.2$ quasar candidates, we 
have discovered a quasar pair separated by 21\arcsec, or about 135\,kpc 
proper. In this paper, we discuss the discovery of this pair, our reasoning 
for why it is a binary quasar (rather than a gravitational lens) and the 
implications of such a pair for quasar clustering at $z\sim5$. All quoted
magnitudes are on the AB system \citep{OkeGunn83} and corrected for Galactic 
extinction using the dust maps of \citet{Sch98}. We adopt a cosmology of 
($\Omega_{\rm m}, \Omega_\Lambda, 
 h\equiv H_0/100\,{\rm km\,s^{-1}\,Mpc^{-1}}) = 
 (0.307,0.693,0.677)$ 
consistent with recent results from {\em Planck} \citep[][]{Planck15}.

\section{Observations}
\label{sec:observations}

\begin{figure}[!t]
\centering
\epsscale{1.15}
\plotone{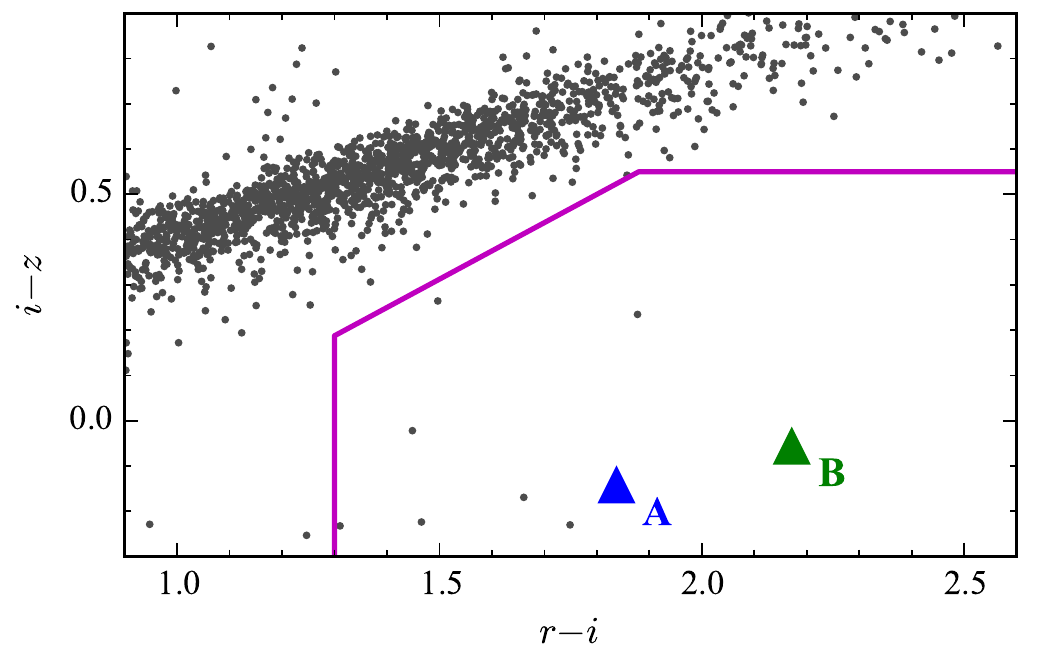}
\caption{Color-color plot displaying the color cuts used to select 
 $z\sim5$ quasar candidates (magenta line). The gray points are objects 
 from the CFHTLS-Wide that pass our morphological and quality cuts and
 that are located in the same 1\,deg$^2$ patch as the quasar pair.
 The blue symbol denotes \qsoa\ and the green symbol \qsob. 
 Error bars are smaller than the symbol size; the colors are clearly 
 different although both are well within our selection boundary.
 \label{fig:colors}} 
\end{figure}

\subsection{Initial Selection from CFHTLS-W1}

In previous work, we measured the $z=5$ quasar luminosity function (QLF) using 
quasars selected from the SDSS Stripe 82 region to a depth of $i=22$ 
\citep{McGreer13}. We are extending this work to fainter quasars using the 
CFHTLS-Wide survey \citep{Gwyn12}. The full CFHTLS-Wide encompasses four fields 
with a total area of 150~deg$^2$ and includes five optical bands, $ugriz$. We 
downloaded the publicly available stacked images\footnote{\cfhtlsurl} and 
generated object catalogs using \sextractor\ \citep{sextractor}. The catalogs 
include PSF photometry derived from the \psfex\ models provided by the CFHTLS. 
We included two of the CFHT-Wide fields in the selection described here; W1 at 
02:18~-07:00 and W3 at 14:18~+54:30.

Full details of our faint $z\sim5$ quasar selection will be provided in a 
future work. Briefly, we use the difference between the elliptical Kron 
aperture magnitude (MAG\_AUTO) flux measurements and the PSF flux measurement 
(MAG\_PSF) from \sextractor\ to obtain a rough star/galaxy separation. Through 
various tests we found that requiring 
${\rm MAG\_AUTO} - {\rm MAG\_PSF} > -0.15$ is highly complete to point sources 
to a limit of $i<23$, while greatly reducing contamination from compact 
galaxies. We further apply a number of quality cuts. First, we require clean 
photometry \sextractor\ FLAGS $<=$ 4. Second, we remove objects lying within 
the masked regions (generally due to bright stars) as provided by the CFHTLS; 
this reduces our effective area by $\sim1$\%. Finally, we remove CFHTLS fields 
for which the stellar locus is poorly matched to a reference locus derived from 
the CFHTLS-Deep survey, indicating issues with the photometric calibration. 
This reduces the areas of both W1 and W3 to 45~deg$^2$ (the full areas are 
72~deg$^2$ and 49~deg$^2$, respectively). 

After applying the morphological and quality cuts and a loose color cut
of $r-i > 0.8$ the resulting density of objects is $\sim$130~deg$^{-2}$
in the two CFHTLS fields. In order to select $z\sim5$ quasar candidates
we adapt the color criteria employed in \citet{McGreer13} to account 
for the bandpass differences between the SDSS and CFHT photometric systems.
This results in the following color cuts:

\begin{align*}
  S/N(u) &< 2.2 \\
  S/N(g) &< 2.2 ~~ {\rm OR} ~~ g-r > 1.8 \\
  r-i &> 1.3 \\
  i-z &< 0.625((r-i) - 1.0) \\
  i-z &< 0.55 \\
\end{align*}

\begin{figure}
\centering
\epsscale{1.15}
\plotone{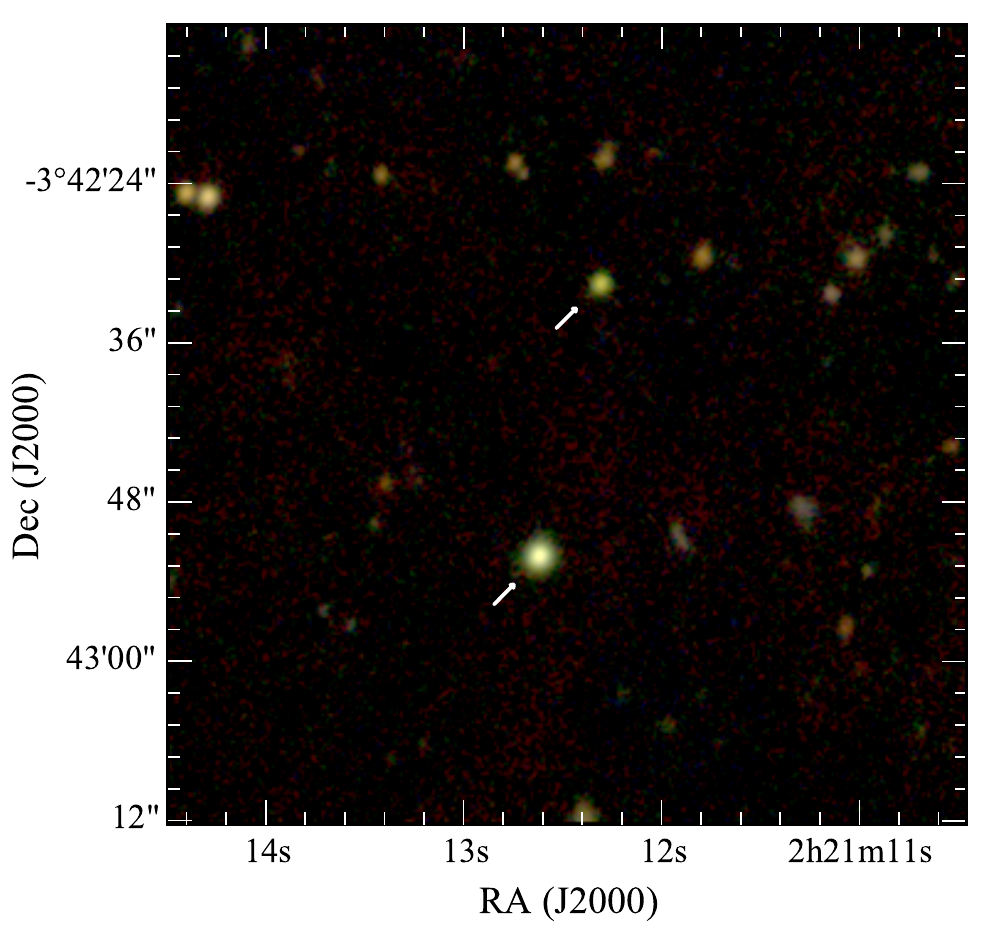}
\caption{Color image of \qsopair\ generated from the CFHTLS $riz$ images. The 
 image is 1\arcmin\ on a side.  The two $z=5$ quasars are indicated with 
 arrows; in this color space they appear green relative to the redder stars 
 and galaxies in the field (note their blue $i-z$ colors in 
 Fig.~\ref{fig:colors}).
 There are no well-detected galaxies between \qsoa\ (lower) and \qsob\ (upper),
 and no apparent overdensity of galaxies in the vicinity. The depths
 of the input images are $g=26.5$, $r=25.9$, and $i=25.6$ \citep{Gwyn12}.
 \label{fig:colorim}} 
\end{figure}

The resulting set of objects were visually examined and those likely to be 
artefacts (e.g., diffraction spikes) were rejected.
In the W1 (W3) field 26 (21) objects are identified as quasar candidates to
a limit of $i=23$. When preparing our observations we noticed
that two of the bright candidates had a very small separation on the 
sky. We examined the imaging and considered both to be viable high
redshift quasar candidates, and thus prioritized them for observation.
However, we emphasize that we did not search for binary candidates
{\it a priori}; rather, we selected the objects simultaneously with
identical criteria.

\begin{figure*}
\centering
\epsscale{1.15}
\plotone{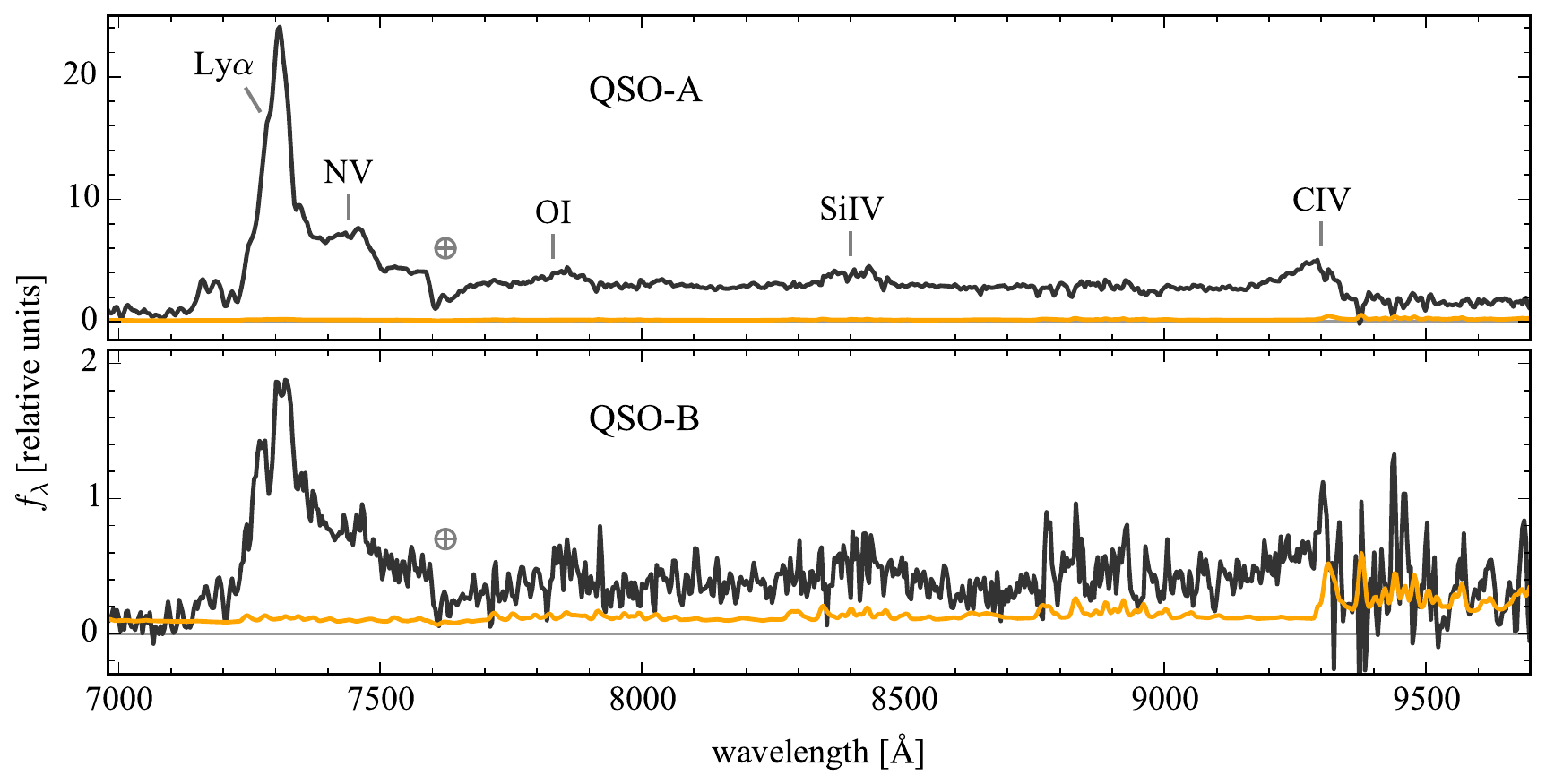}
\caption{MMT Red Channel spectra of \qsoa\ (top) and \qsob\ (bottom).
 The total integration time is 2.9 hrs. The orange lines indicate the
 rms noise level. The absorption feature at $\sim$7620~\AA\ is telluric.
 The locations of prominent quasar emission lines for a redshift of $z=5.0$
 are marked in the upper panel (Ly$\alpha$ is offset for clarity).
 The spectra are remarkably similar except for the
 Ly$\alpha$ emission (see Fig.~\ref{fig:lyacompare}). The red wing of
 the \ion{C}{4} line is strongly affected by night sky emission.
 \label{fig:spectra}} 
\end{figure*}

\subsection{MMT Observations}\label{sec:mmtspec}

We observed \qsopair\ with the Red Channel spectrograph \citep{mmtred} on 
the MMT 6.5m telescope on 2014 Jan 9, 2014 Jan 10, and 2014 Aug 28. All 
observations utilized a 1\arcsec$\times$180\arcsec\ longslit aligned at a 
position angle of $-12.2^{\circ}$ in order to capture both quasar candidates. 
The objects were dispersed with the 270~mm$^{-1}$ grating at a 
resolution of $R\sim640$. For the 2014 Jan 9 observations the central
wavelength was set to 7500~\AA, providing wavelength coverage from
5670\AA\ to 9290\AA, and the total integration time was 70\,min.
For the 2014 Jan 10 and 2014 Aug 28 observations the central wavelength
was 8500\AA~(6600\AA\ $\la \lambda \la$ 1$\mu$m) and the total 
integration times were 45\,min.\ and 60\,min., respectively. In all cases 
the seeing was marginal (1{\farcs}5 -- 2\arcsec) with non-photometric 
conditions.

The spectra were processed in a standard fashion with Pyraf-based
scripts; details of the processing method are given in \citet{McGreer13}.
Wavelength calibration was provided by an internal HeNeAr lamp, and an
approximate flux calibration was obtained from observations of the 
spectrophotometric standard star Feige 110. The calibrations were taken
immediately before the science spectra. The processed spectra from
each of the three nights were interpolated onto a common linear wavelength
grid and combined using inverse-variance weighting. The final spectra
are displayed in Figure~\ref{fig:spectra}.

The spectroscopy immediately confirmed that both candidates are quasars 
at $z\sim5$. In Section~\ref{sec:pairorlens} we interpret the spectra and
other available data in order to determine whether they represent two
quasars at a similar redshift or gravitationally lensed images of a
single source quasar.

We have obtained a total of 19 MMT spectra out of the 47 candidates with
$i<23$ in W1 and W3. A more complete analysis of this sample will be 
presented in a future work. Relevant to this work, we note that {\em all} 
of the observed objects are quasars at $z \ga 4.5$, indicating that the
color selection is highly pure. In addition, our simulations show that
the color selection is highly complete ($>90$\%) in the range 
$4.75 < z \la 5.15$, with a tail to $\sim50\%$ completeness out to $z\sim5.4$ 
\citep[see][for details on the simulation method]{McGreer13}. Although we
have spectra for only 40\% of our candidates, we consider it highly
likely that any similar pair of small-separation quasars would be included
in our target list, and thus we conclude that only one such pair lies within
the 90~deg$^2$ search area to the flux limit of $i<23$\footnote{We searched
our candidate list for additional pairs and found two quasars with a
separation of 80\arcsec; however, they have a redshift difference of
$\Delta{z}=0.15$.}. This estimate of the area of our survey (90~deg$^2$) will 
be used in \S\ref{sec:clustering} to infer the clustering strength of quasars 
at $z\sim5$.

\subsection{Additional Observations}

\qsopair\ lies within the XMM-LSS \citep{xmmlss} survey region, and thus has a 
wide array of multiwavelength observations. Table~\ref{tab:photometry} lists 
photometric observations of the quasar pair, including deep near-IR photometry 
from the UKIDSS Deep eXtragalactic Survey \citep[DXS;][]{ukidss} and deep 
Spitzer photometry from the SWIRE survey \citep{swire}. 
The brighter quasar is also an X-ray source in the XMM-XXL survey and was 
included as an ancillary quasar target in the SDSS-III Baryon Oscillation
Spectroscopic Survey (BOSS) Data Release 12 \citep{DR12}. This ancillary 
program\footnote{\xmmxxlurl} targeted XMM-XXL sources for spectroscopy; 
\qsoa\ was the highest redshift X-ray source in the sample, with a redshift 
of $z=5.011$.

It is important to note that although \qsopair\ happens to lie within a deep 
extragalactic survey field, for our survey it was selected based on optical 
colors from the CFHTLS-Wide alone. 

\begin{figure}
\centering
\epsscale{1.15}
\plotone{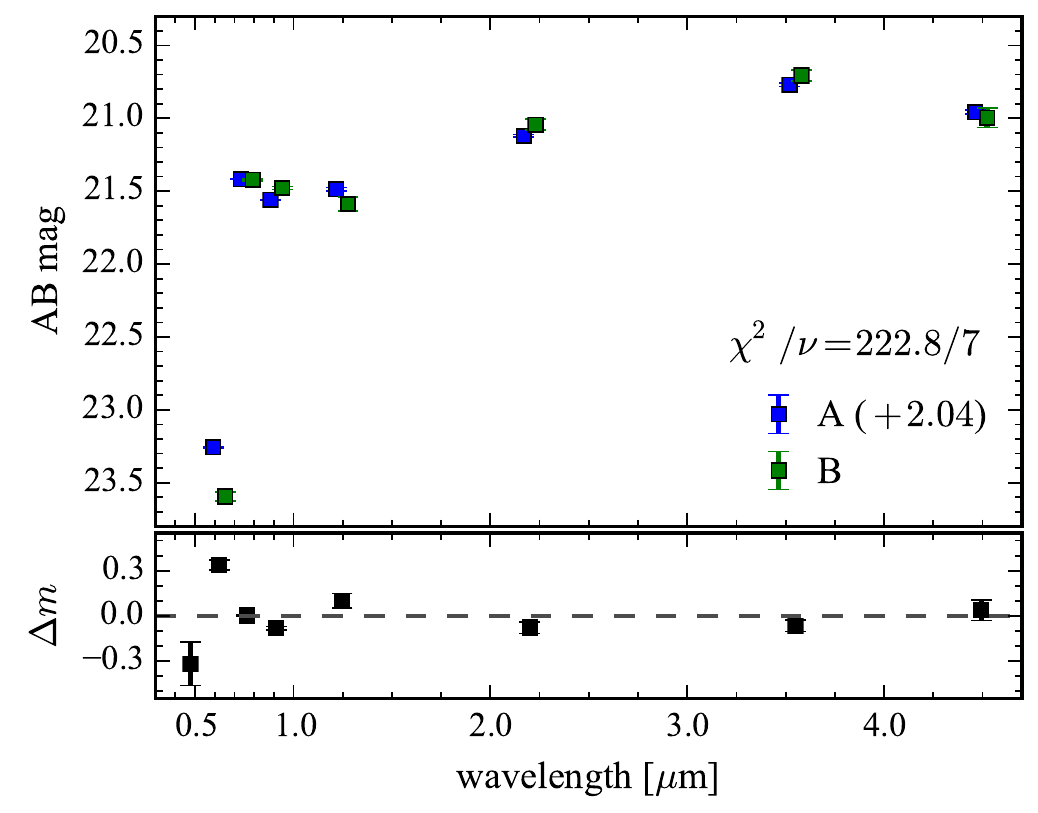}
\caption{SEDs for \qsoa\ and \qsob\ obtained from the CFHTLS-Wide imaging
 ($griz$; the $g$-band measurements are below the plot boundary in the upper
 panel), UKIDSS DXS ($JK$), and Spitzer/SWIRE (3.6$\mu$m and 4.5$\mu$m).
 The upper panel compares the photometric measurements for the two objects;
 the SED for QSO-A is shifted downward so that the $\chi^2$ difference
 between the two SEDs is minimized. The points are 
 offset slightly in wavelength for clarity. The lower panel shows the
 residual magnitude differences after the offset has been applied. There
 are significant differences over the full wavelength range, with the
 largest deviations occuring at the shortest wavelengths, where strong
 rest-UV emission lines contribute significantly to the fluxes. The photometry
 of both objects is always simultaneous due to their proximity, minimizing
 any differences arising from variability. All points include error bars, but 
 they are generally smaller than the symbol size.
 \label{fig:seds}} 
\end{figure}

\begin{deluxetable}{lcc}
 \centering
 \tablecaption{Properties of the binary quasar.}
 \tablewidth{2.25in}
 \tablehead{
  \colhead{} &
  \colhead{QSO-A} &
  \colhead{QSO-B}
 }
 \startdata
 RA (J2000) & 02:21:12.613 & 02:21:12.315 \\
 Dec (J2000) & -03:42:52.19 & -03:42:31.64 \\
$g$     &  $ 24.104 \pm 0.029 $ &  $ 25.822 \pm 0.142 $  \\
$r$     &  $ 21.221 \pm 0.004 $ &  $ 23.594 \pm 0.032 $  \\
$i$     &  $ 19.383 \pm 0.001 $ &  $ 21.423 \pm 0.005 $  \\
$z$     &  $ 19.524 \pm 0.002 $ &  $ 21.478 \pm 0.011 $  \\
$J$     &  $ 19.452 \pm 0.012 $ &  $ 21.588 \pm 0.047 $  \\
$K$     &  $ 19.086 \pm 0.009 $ &  $ 21.044 \pm 0.038 $  \\
$3.6\mu$m &  $ 18.735 \pm 0.012 $ &  $ 20.705 \pm 0.037 $  \\
$4.5\mu$m &  $ 18.923 \pm 0.013 $ &  $ 20.996 \pm 0.067 $  \\
$5.8\mu$m &  $ 19.248 \pm 0.073 $ &  -  \\
$8.0\mu$m &  $ 18.825 \pm 0.046 $ &  -  \\
$24\mu$m &  $ 17.090 \pm 0.051 $ &  - 
 \enddata
\label{tab:photometry}
 \tablecomments{All photometry is on the AB system and corrected for Galactic
  extinction.}
\end{deluxetable}

\section{A Quasar Pair or a Lens?}
\label{sec:pairorlens}

We first consider whether \qsopair\ represents a binary quasar or a pair of 
gravitationally lensed images of a single source at $z=5$. Because lensing 
is achromatic, if they are lensed images the two objects should present 
similar colors at all wavelengths. There are two important effects that can
affect the observed colors: 1) differential reddening along the independent 
light paths to the two images, and 2) time delays between the lensed images
combined with intrinsic source variability. Figure~\ref{fig:seds} presents
the multiwavelength SEDs of \qsopair, where deviations as large as
$\sim$30\% from the mean flux offset are present across a wide range of
wavelengths. These differences are far greater than the photometric 
uncertainties, which are $<1$\% from $g$-band to $K$-band. The statistical
uncertainties are appropriate here since the photometry represents 
{\em relative} flux measurements between the two objects, as the same 
calibrations have been applied to both objects. In addition, because of their 
proximity, the photometry of the two objects is always simultaneous, 
minimizing any differences arising from intrinsic source variability. We apply 
the $\chi^2$ statistic given by  \citet[][their equation 2]{Hen06} to the two 
SEDs as a test of the hypothesis that a simple flux scaling combined with 
photometric scatter accounts for their differences (i.e., they are lensed 
images) and rule this out at high significance, obtaining $\chi^2/\nu$ = 223/7 
from the multiwavelength SEDs\footnote{\citet{Hen06} obtain a median value of 
$\chi^2/\nu$ = 33.1/4 for a sample of SDSS quasars at $2.4 < z < 2.45$; the 
expectation for lensed images in the absence of differential reddening is 
$\chi^2/\nu \sim 1$. To compute this value we use the photometry from $g$ 
through 4.5$\mu$m where both objects are well detected, hence the 7 degrees of 
freedom. If we restrict the data to the $griz$ bands to better compare with 
the SDSS data, we obtain $\chi^2/\nu$ = 212/3.}. Finally, there is no trend in 
the flux differences that would be consistent with reddening.

\begin{figure}
\centering
\epsscale{1.15}
\plotone{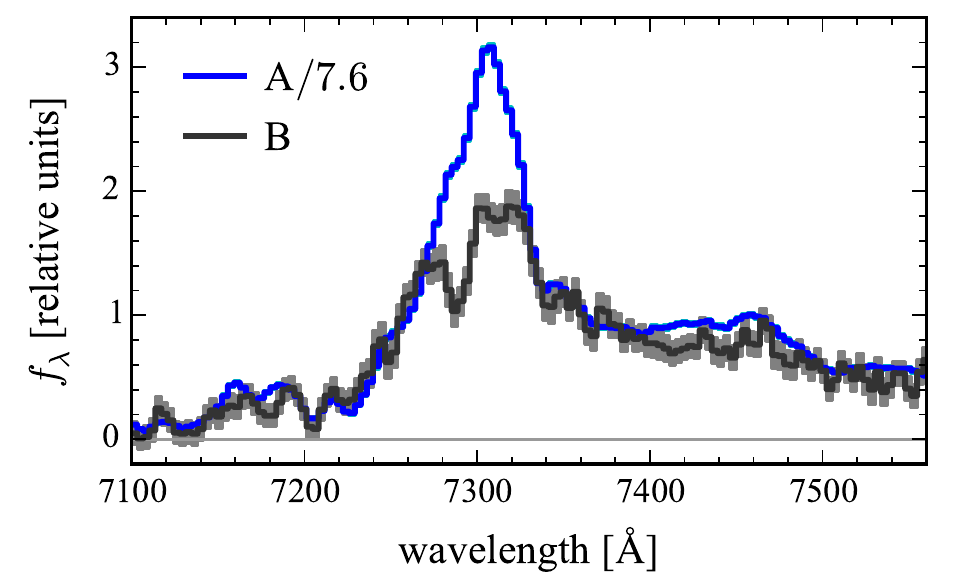}
\caption{Ly$\alpha$ emission line profiles for \qsoa\ (blue) and \qsob\ 
 (dark gray). The rms noise is shown as shaded regions (for \qsoa\ it is
 roughly the same as the line width). The
 \qsoa\ spectrum has been rescaled by the continuum flux ratio between
 the two objects at 8200\AA. The Ly$\alpha$ emission from \qsoa\ has
 larger equivalent width, as does the \ion{N}{5} emission line at
 7440\AA\ (note that the \qsob\ spectrum has a redder continuum slope, 
 consistent with its redder $r-i$ color, see Fig.~\ref{fig:colors}).
 The Ly$\alpha$ emission of \qsob\ has stronger absorption
 features. The wings of the two lines are nearly identical, indicating
 that the velocity offset is extremely small; it is $\la 2$ pixels or
 $\la 270~{\rm km}~{\rm s}^{-1}$.
 \label{fig:lyacompare}} 
\end{figure}

The MMT spectra of the two objects are highly similar to the level of the 
$S/N$ and resolution available. We obtain a small redshift difference from 
fitting the \ion{O}{1} emission line in the MMT spectra; Gaussian fits with 
the IRAF {\it splot} command return $z({\rm A})=5.016$ and $z({\rm B})=5.019$, 
a difference of $\approx160$\kms. However, the \ion{O}{1} line is weakly 
detected in the \qsob\ spectrum and the uncertainty on the line centroid is 
$\approx 300$\kms, thus this difference is not significant. The Ly$\alpha$ 
lines are detected at high $S/N$ in both spectra and the line profiles in the 
wings are nearly identical (Fig.~\ref{fig:lyacompare}), agreeing to within 
two spectral pixels, or $\la270$\kms. As it is difficult to conclusively
state the velocity offset between the two spectra, in the rest of this work
we adopt the difference obtained from the \ion{O}{1} fits, 
$\Delta{z} \la 0.03$.

There are differences between the two spectra that indicate they are not 
likely to originate from a single source quasar. \qsob\ has a clear absorption 
feature just blueward of Ly$\alpha$ at $\sim$7286\AA\ while \qsoa\ has a 
transmission peak at $\sim$7170\AA; both of these features are significantly 
weaker in the opposite spectrum (Figure~\ref{fig:lyacompare}). Also, the 
\ion{N}{5} emission from \qsoa\ is greater than that from \qsob.

Another consideration is that a configuration that produces a large separation 
(21\arcsec) image pair of a $z=5$ quasar is highly unlikely; only three
large-separation lensed quasar systems are known in the entire SDSS 
\citep{Inada03,Dahle13,Rusu13}. Large separations can arise from group- or 
cluster-scale lens masses. For this configuration, the peak in the expected 
lens redshift distribution is at $z\sim0.7$ and the probability of a source 
quasar at $z=5$ with an image separation $>20$\arcsec\ is extremely small 
\citep{Hen07,Li07}. Furthermore, the CFHTLS imaging is sufficiently deep that 
a group-scale overdensity at $z<1$ should be immediately apparent in the 
optical imaging, but as Figure~\ref{fig:colorim} shows, there are no galaxies 
detected between \qsoa\ and \qsob, and no obvious overdensity of galaxies in 
the vicinity.

In conclusion, the mismatched SEDs of the two objects and the lack of any
obvious foreground mass to generate a wide separation lens configuration
strongly argue against the lens hypothesis for this pair of objects. We
proceed to interpret them as a binary quasar.

\section{Implications for Quasar Clustering at $z > 5$}
\label{sec:clustering}

Quasars peak as a luminous population near $z\sim2.5$, and bright quasars at higher redshift are increasingly
rare \citep{Ric06}. Combining the depth and area necessary
to survey large numbers of $z\sim5$ quasars is therefore taxing. Due to the difficulty in studying a significant
quantity of
quasars at high redshift, quasar clustering has only been measured in a statistical fashion out to $z\sim4$, by \citet{She07}, who found
a correlation length of $r_{\rm 0} \sim24$\hMpc. This level of clustering is considered ``large,'' in the sense that
it is at the limits of what might be predicted by theoretical models that use the luminosity function to infer quasar clustering \citep[e.g.][]{Hop07}
and in the sense that clustering measurements at $z\sim2.5$ typically obtain significantly smaller values of 
$r_{\rm 0} \sim8$\hMpc\ \citep[e.g.][]{Whi12,Eft15}. That $r_{\rm 0}$ seems large at
$z\sim4$ motivates further measurements of quasar clustering to determine if $r_{\rm 0}$ remains large
at comparable or higher redshifts.

A small number of close quasar pairs can be used as an alternative method
for quantifying the clustering of high redshift quasars. In the absence
of clustering it would be extraordinarily unlikely to find multiple quasars 
within a small cosmological volume, hence pairs can be used to infer the 
clustering strength required to increase the likelihood of companion quasars
with small separations.
An example of this approach is that of \citet{Sch00}, who used a single binary quasar at $z=4.25$ separated by $\Delta\theta=33.4$\arcsec\ on 
the plane of the sky to infer $r_{\rm 0} \sim12$--30\,Mpc for the correlation length of $z\sim4$ quasars, presaging the 
 \citet{She07} estimate of $r_{\rm 0} \sim24$\hMpc. In our chosen cosmology, the transverse projected separation of the \citet{Sch00}
quasar pair is 160\hkpc\ compared to 90\hkpc\ ($\Delta\theta=21$\arcsec; $z=5.02$) for the pair we have discovered. 

In this section we use our binary quasar to estimate $r_{\rm 0}$ for quasars at $z\sim5$. It may be helpful 
to remember that at $z=5$, an angle of 1\arcsec\ subtends a transverse separation of 26\hkpc\ comoving (4.35\hkpc\ proper).

\subsection{The luminosity function}\label{qlf}

The significance of observing a close pair of quasars relative to random
chance is determined from the QLF. We adopt the recent measurement of the QLF 
at $z=5$ from \citet{McGreer13} based on quasars drawn from the SDSS Stripe 82 
region, extending to a depth of $i=22$. Although \qsopair\ is drawn from a 
deeper survey ($i<23$) and thus requires extrapolation from the 
\citet{McGreer13} results, it is worth noting that both quasars have $i<22$ 
and are thus within the range of the Stripe 82 measurement.

The QLF is typically fit with a double power-law form,
\begin {equation}\label{phi}
\Phi(M,z)=\frac{\Phi^{*}(z)}{10^{0.4(\alpha+1)(M-M^{*})}+10^{0.4(\beta+1)(M-M^{*})}} ~.
\end{equation} 
\citet{McGreer13} estimate the characteristic luminosity to be 
$M^{*}=-27.21$ and the faint and bright end slopes to be $\alpha=-2.03$ and 
$\beta=-4.0$, respectively. The parameter $\Phi^{*}$ is best described by 
a term that evolves with redshift, 
$\log \Phi^{*}(z)=\log \Phi^{*}(z=6)+k(z-6)$, with 
$\log \Phi^{*}(z=6)=-8.94$ and $\rm k=-0.47$ (see \S 6 of \citealt{McGreer13} 
for a detailed discussion on the fitting procedure and redshift evolution of 
the QLF parameters). 

As noted in \S\ref{sec:observations}, our survey of $z\sim5$ quasars has 
progressed such that it is reasonable to assume our binary quasar is drawn 
from a complete survey covering 90\degsq\ to a flux limit of $i < 23$. The 
number density of quasars brighter than $i'=23$ in our survey is calculated 
by taking  the integral of the QLF between the $k$-corrected absolute 
magnitudes,  $M_{i}^{\rm bright}$ and $M_{i'}$, corresponding to the bright 
and faint end of the apparent magnitude range of our survey:
\begin {equation}\label{phist}
n(z,i<i')=\int_{M_{i}^{\rm bright}}^{M_{i'}} dM_{i} ~ \Phi(M_{i},z).
\end{equation} 

We use a constant $k$-correction of $k_{\rm corr}=-2.2$, which is reasonable 
over the full redshift range of interest \citep[see Fig.\ 6 of][]{McGreer13}.
The number densities obtained from the QLF are relatively insensitive to the 
bright limit (in apparent magnitude) adopted for the integration. We ignore
the incompleteness due to our selection efficiency, which would reduce the 
observed number densities. We stress that this makes our measurements more
conservative, in that any incompleteness in our survey would {\em increase}
the inferred clustering signal, as we would be more likely to have missed
additional close pairs. The QLF predicts $\sim0.9$\degsq\ quasars over the 
redshift range $4.7<z<5.2$, which already hints that a quasar pair separated 
by 21\arcsec\ at $z\sim5$ would be highly unusual if quasars were not
significantly clustered at high redshift.

\subsection{Estimating the Correlation Length Using the \citet{Sch00} Formalism}\label{clsres}

Following \citet{Sch00} we determine the correlation length of quasars by 
comparing the single pair we have found to the number of pairs we would 
expect to find in the volume enclosing our pair. The mean number expected in 
a given volume can be determined from the QLF. The odds of finding two quasars 
in that volume (corresponding to our binary quasar) can then be determined 
from the Poisson distribution, as Poisson statistics are an excellent model 
for quasar clustering on small scales where the pairs are independent 
\citep[e.g.][]{Mye06}. The correlation length can be related to the excess 
clustering over random as
\begin{multline}\label{schr}
  \frac{N}{N_{\rm rand}}=
     \frac{\int_{0}^{R} 4\pi r^2 \xi(r)~dr}{\int_{0}^{R}4{\pi}r^2~dr} \\
     = \frac{3}{3-\gamma}\left(\frac{R}{r_0}\right)^{-\gamma}|_{\gamma=2}
     = 3\left(\frac{R}{r_0}\right)^{-2} \ ,
\end{multline}
\noindent where we have adopted a power-law form of $\xi(r)=(r/r_0)^{-\gamma}$ 
with $\gamma=2$ for the slope of the correlation function, as used in many 
studies of the clustering of quasars at high redshift on large and small scales 
\citep[e.g.][]{She10,Whi12,Eft15}.

As it is unclear {\it a priori} to what degree the redshift difference between the 
components of our binary quasar is due to line-of-sight separation versus 
infall, we will calculate ``minimum,'' ``medium'' and ``maximum'' separations 
based on the transverse and line-of-sight comoving separations of our pair 
\citep[again following][]{Sch00}. Our quasar pair is at $z=5.02$, is separated 
by 21\arcsec\ on the plane of the sky, and has a redshift difference of 
$\Delta z \la 0.03$. If the separation of the quasars is entirely in the 
transverse direction, then  the components of our pair are separated by 
810\,kpc comoving (the ``minimum'' separation). If the full redshift 
difference is also due to physical separation, then the components of our pair 
are separated by 16.1\,Mpc comoving (the ``maximum'' separation). If half of 
the redshift difference is attributable to physical separation, then the 
components of our pair are separated by 8.08\,Mpc comoving (the ``medium'' 
separation).

Integrating the \citet{McGreer13} QLF to a limit of $i=23$ over the redshift
range $4.7<z<5.2$ results in a number density of 
$\approx 1.75\times10^{-7}\, \rm Mpc^{-3}$, ignoring selection completeness.
Our ``minimum'' separation of 810\,kpc implies a quasar pair embedded in a
volume of 2.25\,Mpc$^{3}$. Multiplying this by the number density yields an 
expectation of $3.9 \times 10^{-7}$ quasars in the volume of interest. Assuming
a Poisson distribution, the probability of two quasars lying within this 
volume is $7.3\times 10^{-3}$ for an all-sky survey. As our survey only
encompasses 90\degsq, the odds of finding the binary quasar within our survey 
are 1 in 62{,}400, implying that this discovery would have been extremely 
unlikely in the absence of clustering. Substituting 
$N/N_{\rm rand} = 62{,}400$ and $R = 810$\,kpc into Eqn.~\ref{schr} implies 
$r_0 = 117$\,Mpc, or $r_0 = 74$\hMpc.

Similar logic implies $r_0 = 25$\hMpc\ for our ``medium'' separation case 
and $r_0 = 18$\hMpc\ for our ``maximum'' separation case. Thus our 
expectation is that $r_0 \sim 25$\hMpc\ for quasars at $z\sim5$, with a 
lower-bound of $r_0 \sim 18$\hMpc. If we adopt a shallower slope for the 
power law index the correlation length would need to be even greater; e.g.,
for $\gamma=1.8$ the ``medium'' separation case results in 
$r_0 \sim 33$\hMpc. Our measurement implies that the amplitude of quasar 
clustering at $z\sim5$ is similar to that measured at $z\sim4$ by 
\citet{She07}.

\subsection{Estimating the Correlation Length Using the \citet{Hen06} Formalism}

The \citet{Sch00} formalism is simple and straightforward. However, by 
selecting ``minimum'' and ``maximum'' extremes for the distribution of 
peculiar velocities in the redshift-space direction this method ignores 
our expectation for this distribution. In particular, the ``minimum'' case
applies if the two quasars are at the same distance and any redshift 
difference is due to the local velocity field. This is a reasonable 
assumption in our case; however, it is useful to characterize the uncertainty
on that difference, and for that we turn to the method of \citet{Hen06}. This
method accounts for a realistic peculiar velocity distribution for the 
quasars so that we can place a more formal (Poisson) error on the correlation 
length we infer from the existence of the binary.

Following the method described in \citet{Hen06}, we assume that binary 
quasars are well-described by a maximum possible peculiar velocity of 
$|v_{\rm max}| =  2000$\kms. We then project the redshift-space correlation
function over this velocity interval,
\begin{equation}\label{wp}
w_p(R,z)=\int_{-v_{\rm max}/aH(z)}^{v_{\rm max}/aH(z)} \xi_{s}(R, s, z) ds ~,
\end{equation} 
\noindent where $H(z)$ is the expansion rate at redshift $z$ and $\xi_{s}$ 
is the redshift-space quasar correlation function. We include the 
cosmological scale factor $a=1/(1+z)$ to convert distances to comoving 
units.

Given that we are working with a single pair embedded in a relatively 
large volume, $w_p$ could be highly sensitive to changes in the model 
correlation function with scale and/or redshift. Again following 
\citet{Hen06} we ameliorate this effect by measuring the volume-averaged 
correlation function $\bar W_{p}(z)$ over the entire radial bin of 
comoving distance that corresponds to the transverse separation of our 
binary quasar $[R_{\rm min}, R_{\rm max}]$. This results in 
\begin{equation}\label{wpbar}
\bar W_{p}(z) = 
  \frac{\int_{-\frac{v_{\rm max}}{aH(z)}}^{\frac{v_{\rm max}}{aH(z)}}  
  \int_{R_{\rm min}}^{R_{\rm max}} 
    \xi_s(R, s, z)  2\pi R dR ds}{V_{\rm shell}} ~,
\end{equation}
\noindent where $V_{\rm shell}$, the volume of a cylindrical shell in 
redshift space, is given by
\begin{equation}\label{vol}
V_{\rm shell} = \pi \left(R_{\rm max}^{2}-R_{\rm min}^{2}\right) 
                  \left[\frac{2v_{\rm max}}{aH(z)}\right]   ~.
\end{equation} 

Although the redshift-space correlation function is a convolution of the
real-space correlation function with the distribution of peculiar 
velocities, we are projecting over a volume large enough to contain 
the full extent of this distribution function. Hence it is a reasonable
approximation to replace the redshift-space correlation function 
$\xi_{s}(R,s,z)$ with its real-space counterpart $\xi(r,z)$ where 
$\xi(r)=(r/r_{0})^{-\gamma}$ and $r^2 = R^2 + x^2$. We adopt
$\gamma=2$ for the real-space correlation function, as explained in 
\S\ref{clsres}, and assume that this form remains valid for all redshifts 
of interest (i.e.\ $\xi(r,z) = \xi(r)$). The integral in eqn.~\ref{wpbar} 
is instead conducted along the line-of-sight distance $x$,
\begin{multline}\label{wpbarfinal}
\bar W_{p}(R_{\rm min},R_{\rm max},z)= \\
  \frac{\int_{-\frac{v_{\rm max}}{aH(z)}}^{\frac{v_{\rm max}}{aH(z)}} 
         \int_{R_{\rm min}}^{R_{\rm max}}
          \left(\frac{x^2+R^2}{{r_0}^2}\right)^{-\frac{\gamma}{2}} 
            2\pi  R dR dx}
       {V_{\rm shell}} ~.
\end{multline} 

We adopt $[R_{\rm min},R_{\rm max}] = [25, 550]$\hkpc\ (i.e.\ [40, 810]~kpc). 
Here, $R_{\rm max}$ corresponds to the 21\arcsec\ separation of our binary 
quasar at $z=5.02$. We set $R_{\rm min}$ to correspond to 1\arcsec, below
which the seeing in the CFHTLS imaging we used for target selection would 
have precluded the selection of a pair of quasars. The number of expected 
companions of any individual quasar in our survey as a function of transverse 
separation and redshift is then
\begin{multline}\label{nc}
N_{c}  = n(4.7<z<5.2,i<23) \\
  \times V_{\rm shell} ~[1+\bar W_{p}(25, 550, z)] ~,
\end{multline} 
\noindent where $n$ is given by our adopted QLF (see Eqn.~\ref{phist}). By 
varying the correlation length in Eqn.~\ref{wpbarfinal} we obtain a 
range of model values for the number of companions we expect at a separation 
of 21\arcsec\ within our survey volume at $z=5.02$. We then compare the
predicted number of companions to the discovery of a single binary out of 
a sample of 47 quasar candidates,\footnote{We ignore the fact that we do 
not have spectroscopic confirmation for all of our candidates. First, as 
mentioned in \S\ref{sec:mmtspec}, we consider our survey to be highly 
complete to small-separation pairs. Second, our spectroscopy has shown 
that our color selection is highly pure, so that we expect nearly all of 
the 47 candidates to be $z\sim5$ quasars. It is more conservative to use 
the full candidate sample, as including only the objects with spectroscopy 
would greatly {\em increase} the implied clustering signal. In addition, 
the binary reported here was prioritized for observation, so it is 
more correct to adopt the full sample.} i.e., within our sample of 47 
quasars there are two objects within the cylindrical shell defined by 
$V_{\rm shell}$. Thus we are seeking a model for the correlation function 
that results in the expected number of companions to be 
$N_c = 2/47 = 0.04255$.

We find the correlation length that best describes our binary quasar is 
$86$\hMpc, with a 1$\sigma$ lower bound of $25$\hMpc. The lower bound
has been determined using the confidence interval for a single measurement 
provided by \citet{Geh86}.

As with the analysis in \S\ref{clsres}, this result depends on our 
adopted QLF and that our assumed form of the correlation function is valid 
and non-evolving across our redshift range of interest. In particular, in
calculating an $r_0$ that is significantly larger than the scales probed
by our pair, we are implicitly assuming that clustering at small scales 
can be extrapolated to large scales\footnote{It is worth noting that the
projected separation of our binary is just at the scale at which the two-halo
term begins to contribute to the projected correlation function in the HOD
models of \citet{Kayo+12}; see their Fig.~6.} (as was found to be the case 
for low-$z$ quasars by \citealt{Kayo+12} and consistent with results at 
$z\sim3\mbox{--}4$ from \citealt{She10}). Whether we employ the \citet{Hen06} 
formalism or the \citet{Sch00} formalism, we find that the existence of this 
binary quasar implies a correlation length $r_0 > 20$\hMpc, consistent with 
the $r_0 \sim 25$\hMpc\ measured at $z\sim4$ by \citet{She07}. This strongly 
suggests that quasars are at least as clustered at $z\sim5$ as has been found 
at $z\sim4$.


\section{Conclusions}
\label{sec:conclusions}

We have discovered a pair of quasars with apparently identical redshifts of
$z=5.02$ and a separation of 21\arcsec\ on the sky. A number of factors 
argue against the pair being gravitationally lensed images of a single 
source quasar. These include differences in spectral profiles and SED 
shapes, and the fact that no deflector is present between the two quasars 
in relatively deep optical imaging. Assuming the quasar pair is a binary, 
the small projected separation (135 kpc proper) and lack of a clear redshift 
offset implies their physical separation is quite small, within a factor of 
$\sim2\mbox{--}3$ of the virial radius of a typical quasar-hosting dark matter 
halo at high redshift 
\citep[$\sim 10^{12}\mbox{--}10^{13}~M_\sun$,][]{Hop07,She07,Whi12,Eft15}. 
This single detection of a binary at $z=5$ favors models where quasars are 
strongly clustered at high redshift, at least on small scales.

The clustering of quasars is sensitive not only to the triggering 
mechanism(s), but also feedback effects that terminate black hole growth.
Globally, the quasar population experiences a ``downsizing'' trend at
$z\la3$, as activity shifts to lower mass and lower luminosity systems
\citep[e.g.,][]{Ross+13}.
This is often thought to be due to feedback, as the most massive systems
form early but rapidly shut down after their quasar phase, freezing their
black hole mass while the host halos continue to grow. At high redshift
the picture is murkier, with few constraints on the black hole mass and
Eddington ratio distributions. This is demonstrated by \citet{Hop07}, who
compare three disparate models for the continued growth of black holes at
high redshift after their luminous quasar phase. If feedback is efficient
at high redshift, the correlation length should decrease strongly with
increasing redshift. If feedback is inefficient such that the black holes 
grow continuously until $z\sim2$, the correlation length flattens out at 
high redshift. If quasars grow at the same rate as their host halos at $z>3$ 
(the ``maximal'' growth model), the correlation length rises sharply, 
implying that quasars at $z=5$ are more strongly clustered by a factor of a 
few compared to the measurements at $z\sim2.5$.

While repeating the caveat that we have only measured small-scale clustering 
from a single, high-luminosity binary at $z=5$, this observation is most 
consistent with a large correlation length, favoring the models in which 
feedback is highly inefficient. Indeed, \citet{Willott+10} find that the 
Eddington ratios of $z\sim6$ quasars are near unity across a range of 
luminosities, suggesting that fainter quasars are not in a ``decaying'' phase
of black hole growth. 

It is surprising to have found two highly luminous quasars --- presumably 
powered by $>10^8~M_\sun$ black holes and situated in massive dark matter 
halos --- in such close proximity at this redshift. Previous searches have 
relied on wide-area surveys such as the SDSS, whereas we surveyed only 
$\sim0.1$\%\ of the sky and yet discovered a $z=5$ binary quasar {\em bright 
enough to have been selected from SDSS imaging}. Whether this was simply
a chance find will await a more comprehensive search for quasar pairs at
$z \ga 5$.

Measurements of high-redshift quasar clustering on large scales are crucial to 
discriminating between feedback models and better understanding the early 
growth of the most massive black holes in the universe. Such measurements are 
just possible today with wide-area, medium-depth fields such as SDSS Stripe 82 
and the CFHTLS, and ongoing surveys such as the DES, the DESI Imaging Surveys 
(DECaLS, BASS, and MzLS), and KIDS also provide the requisite combination of 
depth and area. Obtaining a fully three-dimensional clustering measurement 
demands a considerable investment in spectroscopic follow-up given the low sky 
density; however, if quasars do cluster strongly at high redshift (as implied 
by our observations), a dense survey over a relatively small area could 
produce a statistically meaningful result.

\vspace{24pt}
IDM and XF acknowledge support from NSF grants 11-06682 and NSF 15-15115.
ADM and SE were supported in part by NASA ADAP award NNX12AE38G and by NSF 
awards 12-11112 and 15-15404.
This research made use of Astropy, a community-developed core Python package 
for Astronomy \citep{astropy}.
Observations reported here were obtained at the MMT Observatory, a joint 
facility of the Smithsonian Institution and the University of Arizona.
Also based on observations obtained with MegaPrime/MegaCam, a joint project of CFHT and CEA/DAPNIA, at the Canada-France-Hawaii Telescope (CFHT) which is operated by the National Research Council (NRC) of Canada, the Institut National des Science de l'Univers of the Centre National de la Recherche Scientifique (CNRS) of France, and the University of Hawaii. The observations at the Canada-France-Hawaii Telescope were performed with care and respect from the summit of Maunakea which is a significant cultural and historic site. This work is based in part on data products produced at the Canadian Astronomy Data Centre as part of the Canada-France-Hawaii Telescope Legacy Survey, a collaborative project of NRC and CNRS.

{\it Facilities:} 
 \facility{MMT (Red Channel spectrograph)}, 
 \facility{CFHT (MegaCam)}

\bibliographystyle{apj}

\end{document}